\documentstyle[amssymb,thmsa,graphicx]{article}

\input{tcilatex}
\begin{document}

\title{Precision microwave dielectric and magnetic susceptibility measurements of
correlated electronic materials using superconducting cavities }
\author{Z. Zhai, C. Kusko, N. Hakim and S. Sridhar \\
Physics Department, Northeastern University, 360 Huntington Avenue, \\
Boston, MA 02115}
\maketitle

\begin{abstract}
We analyze microwave cavity perturbation methods, and show that the
technique is an excellent, precision method to study the dynamic magnetic
and dielectric response in the $GHz$ frequency range. Using superconducting
cavities, we obtain exceptionally high precision and sensitivity for
measurements of relative changes. A dynamic electromagnetic susceptibility $%
\tilde{\zeta}(T)=\zeta ^{\prime }+i\zeta ^{\prime \prime }$ is introduced,
which is obtained from the measured parameters: the shift of cavity resonant
frequency $\delta f$ and quality factor $Q$. We focus on the case of a
spherical sample placed at the center of a cylindrical cavity resonant in
the $TE_{011}$ mode. Depending on the sample characteristics, the magnetic
permeability $\tilde{\mu}$, the dielectric permittivity $\tilde{\varepsilon}$
and the complex conductivity $\tilde{\sigma}$ can be extracted from $\tilde{%
\zeta}_{H}$.

A full spherical wave analysis of the cavity perturbation indicates that :
(i) In highly insulating samples with dielectric constant $\varepsilon
^{\prime }\sim 1$, the measured $\tilde{\zeta}_{H}\approx \tilde{\chi}_{M}$,
enabling direct measurement of the magnetic susceptibility. The sensitivity
of the method equals or surpasses that of dc SQUID measurements for the
relative changes in magnetic susceptibility. (ii) For moderate $\tilde{%
\varepsilon}$ and conductivity $\sigma $, $\tilde{\zeta}_{H}\varpropto 
\tilde{\varepsilon}+i\omega \tilde{\sigma}/\varepsilon _{o}-1$, thus
enabling direct measurement of the sample dielectric constant $\tilde{%
\varepsilon}$, even though the sample is placed in a microwave magnetic
field. (iii) For large $\sigma $ we recover the surface impedance limit.
(iv) Expressions are provided for the general case of a lossy dielectric
represented by $\tilde{\varepsilon}+i\omega \tilde{\sigma}/\varepsilon _{o}$%
. We show that an inversion procedure can be used to obtain $\tilde{%
\varepsilon}+i\omega \tilde{\sigma}/\varepsilon _{o}$ in a wide range of
parameter values.

This analysis has led to the observation of new phenomena in novel low
dimensional materials. We discuss results on magneto-dynamics of the 3-D
antiferromagnetic state of spin chain compound $Sr_{2}CuO_{3}$. In
dielectric susceptibility measurements in $Sr_{14}Cu_{24}O_{41}$, we
directly observe a dielectric loss peak. Dimensional resonances in the
paraelectric material $SrTiO_{3}$ are shown to occur due to the rapid
increase of dielectric constant with decreasing temperature. The cavity
perturbation methods are thus an extremely sensitive probe of charge and
spin dynamics in electronic materials.
\end{abstract}

\section{Introduction}

The continuing discovery of new electronic materials calls for new methods
of measuring their electric and magnetic properties. Microwave cavity
perturbation techniques have proved to be very useful for the study of
transport dynamics at microwave frequencies\cite
{Sridhar88,Gruner99,Brodwin65,Khanna74}, in materials such as
semiconductors, magnetic ferrites and exotic materials such as Charge and
Spin Density Waves \cite{Ong77}.

In all of these previous studies normal metal cavities were used. To study
the (then) newly discovered high temperature superconductors (HTS), the use
of superconducting cavities was introduced by Sridhar and Kennedy\cite
{Sridhar88}. The reduction in background absorption by a factor of $10^{4}$
from a normal metal cavity enabled the measurement of absorption in small,
single crystal superconductors and thin films. The surface impedance $\tilde{%
Z}_{s}=R_{s}-iX_{s}$ was obtained in terms of changes of the cavity
parameters : the shift in frequency $\delta f$ and quality factor $Q$.
Subsequently the concept of the ``hot finger'' technique introduced in \cite
{Sridhar88} has been used in measurements in other laboratories also with
the purpose of studying HTS \cite{Bonn,Wu}.

In this paper, we present a reanalysis of the cavity perturbation technique,
and describe a new application utilizing superconducting microwave cavities,
to study dynamic electric and magnetic susceptibilities of strongly
correlated electronic materials. We focus on the configuration where the
sample is placed at a microwave magnetic field maximum of the $TE_{011}$
mode.

\begin{enumerate}
\item  We introduce an electromagnetic susceptibility $\tilde{\zeta}=\zeta
^{\prime }+i\zeta ^{\prime \prime }$, which provides a useful framework to
discuss the results of the microwave measurements. We use $\tilde{\zeta}_{H}$
to note the case where the sample is measured in a microwave magnetic field
(e.g. in the $TE_{011}$ mode), and $\tilde{\zeta}_{E}$ when the sample is
placed in a microwave electric field (e.g. in the $TM_{010}$ mode).
Depending on sample properties, the measured parameter $\tilde{\zeta}$ can
be related to the sample magnetic permeability ($\tilde{\mu}=1+\tilde{\chi}%
_{M}$) and dielectric permittivity ($\tilde{\varepsilon}=1+\tilde{\chi}_{P}$%
), where $\tilde{\chi}_{M}(\tilde{\chi}_{E})$ are the magnetic(electric)
susceptibilities, the conductivity $\tilde{\sigma}$ and the surface
impedance $Z_{s}$. These various limits are discussed in detail in the paper.

\item  For highly insulating samples with $\tilde{\varepsilon}\sim 1$, the
technique is a very sensitive method of measuring the magnetic
susceptibility, since $\tilde{\zeta}\sim \tilde{\chi}_{M}=\chi _{M}^{\prime
}+i\chi _{M}^{\prime \prime }$. The sensitivity of this technique is
compared with others, and it is shown that the microwave method, when
superconducting cavities are used, can equal or even exceed that of a dc
SQUID for relative changes in susceptibility, such as with changing $T$. It
also yields results on samples (typically mm-sized) in which comparable ac
susceptibility measurements do not have sufficient sensitivity. As an
example of this technique we show that it yields information on
magnetodynamics in a spin chain material $Sr_{2}CuO_{3}$.

\item  When the sample conductivity or dielectric constant is substantial,
the measurements are dominated by these parameters. For insulating samples
with even moderate dielectric constants $\varepsilon ^{\prime }$, the
experiments are a direct measurement of $\tilde{\varepsilon}=\varepsilon
^{\prime }+i\varepsilon ^{\prime \prime }$. {\em Thus we are able to measure 
}$\tilde{\varepsilon}${\em \ even though the sample is placed in a microwave
magnetic field maximum. }(In fact the $H_{\omega }$ field measurements of $%
\tilde{\varepsilon}$ have an advantage over $E_{\omega }$ measurements as
they are not subject to the so-called depolarization peak). We describe an
inversion procedure to obtain the complex dielectric constant $\tilde{%
\varepsilon}+i\tilde{\sigma}/\omega \varepsilon _{o}$ from the measured
data. A spectacular example of the dielectric measurements is the
observation of a dielectric loss peak in $\varepsilon ^{\prime \prime }(T)$
due to dielectric relaxation in the spin ladder compound $%
Sr_{14}Cu_{24}O_{41}$.

\item  For sufficiently large $\varepsilon ^{\prime }$, dimensional
resonances can occur when the microwave essentially enter into the sample.
An striking example of this is presented in data on $SrTiO_{3}$.

\item  When the conductivity $\sigma $ is appreciable, it can lead to an
eddy current contribution resulting in a peak in absorption with increasing
conductivity. (This is the magnetic analog of the so-called depolarization
peak for $E_{\omega }$ field measurements). For large conductivity the
results tend to the surface impedance limit. This is the limit used in
previous measurements of the surface impedance of metals and
superconductors. This paper presents a unified approach which encompasses
both the insulating and highly metallic limits.
\end{enumerate}

The cavity perturbation method discussed here yields unique information on
spin and charge dynamics at short time scales between Neutron Scattering and
NMR and $\mu SR$, and has led to the observation of some unique phenomena in
quantum magnets, dielectrics and superconductors.

\section{Description of apparatus and measurement technique}

A right cylindrical cavity (inner radius $7/8$ $inch$ and axial length $1$ $%
inch$) was made of pure Niobium ($Nb$), which is a superconductor below $%
T_{c}=9.2K$. The cavity was fabricated in three pieces: two end plates with
the needed holes and one center ring. The top plate has a center pumping
hole ($3.56$ $mm$ diameter), and two coupling holes ($3.56$ $mm$ diameter),
the bottom plate has one centrally located hole ($6.7$ $mm$ diameter),
through which the sample is inserted into the cavity. The $TE_{011}$ mode is
degenerate with the $TM_{111}$ mode. As $TE_{011}$ is the desired operating
mode, the diameter of these coupling holes was chosen to provide enough
perturbation to split the two modes more than $40MHz$ apart. The high
quality $Nb$ stock was carefully machined at very low speed to the needed
shape and then polished without lubricant, which would otherwise cause
oxidation on the $Nb$ surface. Each piece was then annealed, and the grains,
which grew due to annealing, vary from sub millimeter size to roughly $4$ $%
mm $ diameter. The three-piece cavity was tightly held by a stainless steel
assembly consisting of a top ring, a center piece for alignment and a bottom
ring. The whole resonator was then mounted in an alignment frame, supported
on the top by a stainless steel dewar probe ($10.16$ $cm$ diameter and $1$ $%
m $ long) and, on the bottom, with a sealed copper cup ($10.16$ $cm$
diameter and $10.16$ $cm$ long) with a removable bottom copper plate. Indium
seals were used so that the entire assembly were vacuum tight.
Superconducting operation of the cavity was accomplished using a bath of
liquid $_{4}He$.

\begin{figure}[tbph]
\begin{center}
\includegraphics*[width=0.45\textwidth]{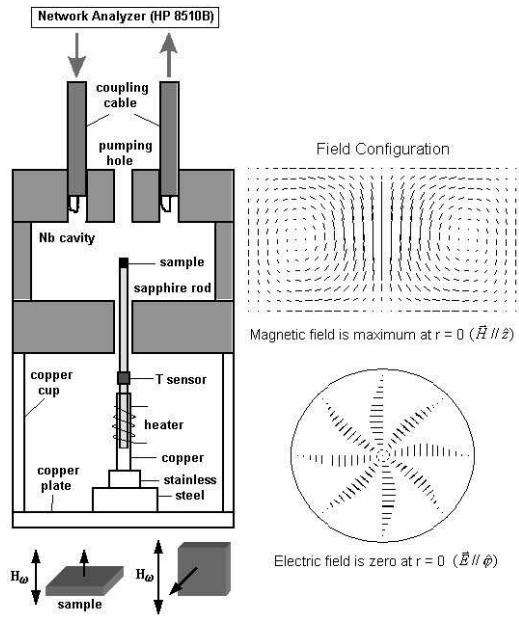}
\end{center}
\par
\caption{(Left) Diagram for 10GHz superconducting cavity resonant, showing
the sample location and the hot finger arrangement to vary sample
temperature, while keeping the cavity walls at the bath temperature. By
varying sample orientation with respect to $H_{\omega }$ as shown, the
anisotropic response can also be measured. (Right) Field configuration in
the $TE_{011}$ mode. Also shown is the (top) $H_{\omega }$ field lines in
the ($r,z$) plane plane, and (bottom) $E_{\omega }$ fields at the central ($%
r,\theta ,z=L/2$) plane of the cavity. For this mode the sample is located
at the $H_{\omega }$ field maximum, and at $E_{\omega }=0$.}
\label{exptfig}
\end{figure}

A small piece of sample was mounted on the top of a sapphire rod ($1.56mm$
diameter $\times $ $52mm$ long) using very little Apiezon-N grease. The
anisotropic response of the sample can be measured by mounting the sample in
different orientations with respect to the applied microwave field for a
given mode, as shown in the set-up diagram (Fig. \ref{exptfig}). The
sapphire rod with sample was inserted into the cavity, along its axis from
the bottom, such that the sample is stationed exactly at the center of the
cavity. Support and adjustment of the sapphire rod was provided by a copper
tube ($20mm$ long), the overlap between copper tube and sapphire rod is
adjustable and finally fixed with GE-Varnish to guarantee good thermal
contact. The copper tube was brazed at the end to a $6.35$ $mm$ diameter
stainless tube (wall thickness $0.15mm$), and the stainless tube was brazed
to the bottom copper plate to form a thermal path to the bottom plate which
is in contact with liquid $_{4}He$.

To heat the sample to higher temperature, a $50$ $\Omega $ heating coil ($%
0.1 $ $mm$ Nichrome wire of $6.5$ $\Omega /ft.$) was wound around the copper
tube, and the control of the sample temperature was accomplished using an
external temperature controller (Lake Shore DRC 82C), with a Silicon Diode
temperature sensor (Lake Shore DT 470) which is attached to the sapphire rod
outside the cavity. Another temperature sensor is put in the Helium chamber
to monitor the bath temperature.

Microwaves were generated with a HP8510B network analyzer, a HP8341B
synthesized sweeper and a HP8516A reflection/transmission test set, and
coupled into and out of the resonator from the top, through two adjustable $%
50\Omega $ coaxial lines, each terminated in a loop. One very useful feature
of the design is the ability to vary the coupling to the resonator by moving
the lines in and out along the axis of the resonator. Thus it is possible to
achieve critical coupling and weak coupling over a wide range of the
resonator quality factor $Q$ ($10^{4}$ $-$ $10^{8}$). For fixed coupling,
input microwave power can be easily varied and the nonlinear effect of the
some samples can be observed\cite{Jacobs96,Zhai97}. The resonant cavity
operated at desired $TE_{011}$ mode has the highest quality factor $Q$ about 
$2\times 10^{8}$ at $2K$ bath temperature.

\section{Electrodynamic basis of the measurement}

A small sample of volume $V_{s}$ placed in a resonant cavity causes the
resonant frequency $f$ and quality $Q$ factor to change by a small amount $%
\delta f$. Assuming the shift in frequency is much smaller than the resonant
frequency, $\delta f\ll f$, the change in cavity parameters can be expressed
as\cite{Brodwin65,Muller39,Slater46,Peligrad98} 
\begin{equation}
-\frac{\delta \tilde{f}}{f}\simeq \frac{(\tilde{\mu}-1)\mu _{o}}{4\langle
U\rangle }\int \vec{H}\cdot \vec{H}_{o}dV_{s}+\frac{(\tilde{\varepsilon}%
-1)\varepsilon _{o}}{4\langle U\rangle }\int \vec{E}\cdot \vec{E}_{o}dV_{s}
\label{cavperb1}
\end{equation}
where the complex frequency shift $\delta \tilde{f}=\delta f-i\Delta f$. $%
\delta f\equiv f_{s}-f_{c}$ and $\Delta f\equiv \Delta f_{s}-\Delta f_{c}$
are the changes in the resonant frequency $f$ and the resonance width $%
\Delta f$ respectively with (subscript $s$) and without (subscript $c$) the
sample. The resonance width is related to the cavity $Q$ factor by $\Delta
f=f/2Q$. $\langle U\rangle $ is the energy stored in the cavity of the
resonant mode. $(\vec{H}_{o},\vec{E}_{o})$ and $(\vec{H},\vec{E})$ are the
cavity field configurations before and after the sample perturbation. A time
dependence $e^{-i\omega t}$ is assumed. $\tilde{\varepsilon}$ and $\tilde{\mu%
}$ are the complex permittivity and permeability. We define the magnetic
susceptibility as $\tilde{\chi}_{M}=\tilde{\mu}-1=\chi _{M}^{\prime }+i\chi
_{M}^{\prime \prime }$ and the dielectric susceptibility as $\tilde{\chi}%
_{P}=\tilde{\varepsilon}-1=\chi _{P}^{\prime }+i\chi _{P}^{\prime \prime }$.

It is convenient to discuss experimental results in terms of an effective
dynamic or electromagnetic susceptibility $\zeta $, 
\begin{equation}
\delta \tilde{f}\equiv -g\tilde{\zeta}\equiv -g(\zeta ^{\prime }+i\zeta
^{\prime \prime })  \label{zetadefine}
\end{equation}
where $g$ is a sample geometrical factor, which is specific to the mode
geometry and sample shape. Under appropriate conditions, $\tilde{\zeta}$ can
be directly associated with the conventional magnetic $\tilde{\chi}_{M}$ or
dielectric $\tilde{\chi}_{P}$ susceptibilities, as will be shown below.

To proceed further requires additional assumptions. Various approximations
have been made, called the ``Quasistatic'' (QS), ``Extended Quasistatic''
(EQS) and Spherical Wave (SW) analysis, depending on the approximation used
to obtain the fields $(\vec{H},\vec{E})$. An extensive analysis was carried
out by Brodwin and Parsons (BP) \cite{Brodwin65}, which covers essentially
all the regimes needed for the experimental measurement discussed here. In
the following we use BP and analyze the various regimes.

\section{Spherical sample in $TE_{011}$ mode}

The $TE_{011}$ configuration is well suited as a probe of the microwave
response of materials because of the very high $Q$'s achievable in this
mode. In the present experiments, the sample is located at the center of the
cavity on the axis. In this location we have maximum uniform axial magnetic
field $\vec{H}$ and zero electric field $\vec{E}$. (See Fig.\ref{exptfig}
for spatial profiles of the $\vec{H}$ and $\vec{E}$ fields). In the
following we use the analysis of BP, details of which are given in the
appendix.

The geometrical factor $g$ of a spherical sample is given as 
\begin{equation}
g=\frac{f}{J_{0}^{2}(\beta _{01}^{^{\prime }}r_{o})\left[ 1+\left( \frac{\pi 
}{L\beta _{01}^{\prime }}\right) ^{2}\right] }\cdot \frac{V_{s}}{V_{c}}
\label{gforte011}
\end{equation}
where $V_{c}$ is the volume of the empty cavity. $\beta _{01}^{^{\prime
}}r_{o}$ is the first root of Bessel function $J_{0}^{^{\prime }}(\beta
r_{o})=0$. Using the cavity inner radius $r_{o}=7/8$ $inch$, and axial
length $L=1$ $inch$, we get $g\approx $ $1.036\times 10^{15}\cdot V_{s}$ $[$m%
$^{-3}\cdot $sec.$^{-1}]$, where $V_{s}$ is the sample volume.

The important parameters that define the analysis are the wave vector inside
and outside the sample : $k_{o}=\omega /c$ and $k=k_{o}\sqrt{\tilde{%
\varepsilon}+i\tilde{\sigma}/\omega \varepsilon _{o}}$. The full-wave
analysis yields in principle (see Appendix A), results of the frequency
shift due to sample perturbation for a large range of sample sizes and
material properties. However in all cases of experimental interest, the
sample size is much smaller than the cavity dimensions, so that the
condition $k_{o}a\ll 1$ is rigorously satisfied. For example, if $a=1mm$ and
the measuring frequency is $10GHz$, then $k_{o}a\simeq 0.2$. In this limit,
we obtain 
\begin{equation}
\tilde{\zeta}_{H}=\frac{3}{2}\left( \frac{(2\tilde{\mu}+1)j_{1}(ka)-\sin (ka)%
}{(\tilde{\mu}-1)j_{1}(ka)+\sin (ka)}\right)  \label{genlzeta}
\end{equation}
We use the subscript $H$ to denote that the $EM$ susceptibility $\tilde{\zeta%
}$ is being measured with an applied microwave magnetic field $\vec{H}%
_{\omega }$.

This general form is in principle valid for arbitrary $ka$ which is
determined by material properties $\tilde{\mu}$, $\tilde{\varepsilon}$ and $%
\tilde{\sigma}$. However in this form it is not very useful. It is therefore
necessary to consider the different limits of this expression. Below we
discuss the various limits and their applicability.

\subsection{Magnetic permeability and susceptibility measurements}

More generally the result in this limit can be written as : 
\begin{equation}
\tilde{\zeta}_{H}=3\frac{\tilde{\mu}-1}{\tilde{\mu}+2}+\frac{9}{10}\left[ 
\frac{\tilde{\mu}^{2}-6\tilde{\mu}+4}{\left( \tilde{\mu}+2\right) ^{2}}%
\left( k_{o}a\right) ^{2}+\frac{\tilde{\mu}}{\left( \tilde{\mu}+2\right) ^{2}%
}\left( ka\right) ^{2}\right] .  \label{mumeasure}
\end{equation}
Clearly the experiment measures $\tilde{\mu}$ only if the second term is
negligible. This may be possible in ferromagnetic samples where $\mu
^{\prime }\gg 1$ provided the spins continue to respond at microwave
frequencies. For weakly paramagnetic samples, we have 
\begin{equation}
\tilde{\zeta}_{H}=\tilde{\chi}_{M}\,\;\text{;\thinspace \thinspace
\thinspace \thinspace \thinspace \thinspace \thinspace \thinspace \thinspace
\thinspace \thinspace \thinspace \thinspace \thinspace \thinspace \thinspace
\thinspace \thinspace \thinspace \thinspace \thinspace \thinspace \thinspace
\thinspace \thinspace \thinspace \thinspace \thinspace \thinspace \thinspace
\thinspace \thinspace \thinspace \thinspace \thinspace \thinspace \thinspace 
}\;(k_{o}a)^{2}\tilde{\chi}_{P}\ll \tilde{\chi}_{M}\text{ }
\label{mumeasure2}
\end{equation}
This limit is only achieved provided the sample is highly insulating and the
dielectric constant is nearly $1$.

\subsubsection{Sensitivity and accuracy of magnetic susceptibility
measurements}

Having established the relationship between magnetic susceptibility $\tilde{%
\chi}_{M}=\chi _{M}^{\prime }+i\chi _{M}^{\prime \prime }$ and measured
electromagnetic susceptibility $\tilde{\zeta}_{H}$ in Eq. \ref{mumeasure2},
we can estimate the measurement sensitivity of the technique. Clearly the
sensitivity is associated with both the size of samples and the cavity
resonant frequency $f$. The bigger the sample size is, the higher the
sensitivity is, as seen from Eq.\ref{zetadefine}, \ref{gforte011}, \ref
{mumeasure2}, provided we still retain the small perturbation limit.
Assuming a typical small sample has the dimension of $V_{s}\thicksim 1\times
1\times 0.5mm^{3}$, as in our experiment, we can detect the frequency shift $%
\delta f$ and the absorption width $\Delta f$ as small as $1Hz$ in a
resonant frequency of $10^{10}Hz$. This results in a sensitivity limit of $%
\delta \zeta _{H}^{\prime }\thicksim 10^{-6}$ and hence $\delta \chi
_{M}^{\prime }\thicksim 10^{-6}$. For comparison, Table 1 lists the
sensitivities of some commonly used techniques for magnetic susceptibility $%
\chi $ measurements\cite{Heinbook}. In these measurements, $\chi $ generally
has the form \cite{Heinbook} 
\begin{equation}
\chi \thickapprox \frac{M}{V_{s}H}  \label{dcsusc}
\end{equation}
where $M$ is the magnetic moment in [$A\cdot m^{2}$], $H$ is the applied
magnetic field in [$A\cdot m^{-1}$]. If the same sample with volume $V_{s}$
is used for all these measurements, assuming an applied field of $H\simeq
10^{5}A/m$ corresponding to typical microwave fields, we can compare their
sensitivities, as listed in Table 1. Note that Eq. \ref{mumeasure2} gives
the magnetic volume susceptibility $\tilde{\chi}_{M}=dM/dH$ in unit of SI or
MKS (dimensionless). In CGS, it is usually expressed in [$emu\cdot cm^{-3}$%
]. To convert from CGS to SI, a conversion multiplying factor of $4\pi $ is
used.

\begin{table}[tbh]
\begin{center}
\begin{tabular}{|l|l|l|c|}
\hline
Method & $M$ $[A\cdot m^{2}]$ & Accuracy in [$m^{3}/V_{s}$] & $\chi $ value
\\ \hline
Superconducting Cavity & \multicolumn{1}{|c|}{$-$} & \multicolumn{1}{|c|}{$%
10^{-16}$ (in $f\simeq 10GHz$)} & relative \\ \hline
dc SQUID & \multicolumn{1}{|c|}{$10^{-11}$} & \multicolumn{1}{|c|}{$10^{-16}$%
} & absolute \\ \hline
ac - $\chi $ & \multicolumn{1}{|c|}{$5\times 10^{-10}$} & 
\multicolumn{1}{|c|}{$5\times 10^{-15}$} & absolute \\ \hline
Vibrating-sample magnetometer & \multicolumn{1}{|c|}{$5\times 10^{-8}$} & 
\multicolumn{1}{|c|}{$5\times 10^{-13}$} & absolute \\ \hline
Alternating-Gradient-Force Magnetometer & \multicolumn{1}{|c|}{$10^{-11}$} & 
\multicolumn{1}{|c|}{$10^{-16}$} & absolute \\ \hline
\end{tabular}
\end{center}
\caption{The comparison of magnetic susceptibility measurements for
microwave cavity technique and other techniques.}
\label{mictab1}
\end{table}

The table shows that the hot finger cavity perturbation technique
undoubtedly has one of the highest measurement sensitivities available.
While other methods may require relatively large sample size and large
applied field $H$, these are not required in the microwave measurements.
However this high sensitivity is achieved only for relative changes, such as
for instance with varying temperature. The precision for absolute
measurements is much less due to small uncertainties in sample location.

\subsection{Lossy Dielectric, Permittivity and Surface Impedance Measurements
}

For even moderate conductivity and dielectric constants, the magnetic
contribution is overwhelmed by the dielectric and conductivity
contributions. Taking $\tilde{\mu}\sim 1$, in the limit $\tilde{\chi}%
_{M}\,\ll (k_{o}a)^{2}\tilde{\chi}_{P}$ , we have : 
\begin{equation}
\tilde{\zeta}_{H}=-\frac{3}{2}\left( 1-\frac{3}{(ka)^{2}}+\frac{3\cot ka}{ka}%
\right) \,\,\,\,  \label{lossydielec}
\end{equation}

\subsubsection{Dielectric permittivity and susceptibility measurements}

The small $ka$ limit of this results leads directly to a measurement of the
dielectric permittivity or susceptibility: 
\begin{eqnarray}
\tilde{\zeta}_{H} &\approx &\frac{1}{10}(k_{o}a)^{2}(\tilde{\varepsilon}+i%
\tilde{\sigma}/\omega \varepsilon
_{o}-1)\,\,\,;\,\,\,\,\,\,\,\,\,\,\,\,\,\,\,\,\,\,\,\,\tilde{\chi}_{M}\,\ll
(k_{o}a)^{2}\tilde{\chi}_{P}\text{ , } \,\,ka\ll 1  \label{elecperm} \\
&\approx &\frac{1}{10}(k_{o}a)^{2}\tilde{\chi}_{P}\,\,\,;\,\,\,\,\,\,\,\,\,%
\,\,\,\,\,\,\,\,\,\,\,\,\,\,\,\,\,\,\,\,\,\,\,\,\,\,\,\,\,\,\,\,\,\,\,\,\,\,%
\,\,\,\,\,\,\,\,\,\tilde{\sigma}=0,\,\tilde{\chi}_{M}\,\ll (k_{o}a)^{2}%
\tilde{\chi}_{P}\text{ , }\,\, ka\ll 1  \nonumber
\end{eqnarray}

A surprising conclusion is that {\em one can measure the dielectric
properties even though the sample is placed in a pure microwave magnetic
field}. We emphasize that this conclusion has nothing to do with the spatial
variation of the $E$-field near the cavity axis. It simply arises from the
wave equation and {\em holds, within geometric factors, even in a homogenous
magnetic field and with zero electric field}, such as can be achieved in a
split ring resonator \cite{Bonn}.

This method of measuring $\tilde{\varepsilon}$ has one important advantage
over $E$-field cavity perturbation measurements. The measured quantity is
directly proportional to $\tilde{\varepsilon}-1$ and holds even when $\tilde{%
\varepsilon}\gg 1$ so long as $(k_{o}a)^{2}(\tilde{\varepsilon}-1)<1$, while
in the $E$-field method the measured frequency shifts are proportional to $(%
\tilde{\varepsilon}-1)/(\tilde{\varepsilon}+2)$, due to so-called
depolarization effects (see Appendix B), and can obscure the direct
interpretation of the results.

For general $\tilde{\varepsilon}$ Eq.\ref{lossydielec} can be inverted to
obtain $\tilde{\varepsilon}$ from the measured $\tilde{\zeta}_{H}$. Examples
of such inversions are presented later.

Note that the dielectric permittivity measured is that appropriate to the
plane perpendicular to the direction of the magnetic field $\vec{H}_{\omega
} $. This is the direction of the displacement currents, and also the
induced conduction currents. If the response in the plane is anisotropic
then the measured $\tilde{\varepsilon}$ will be an appropriate mixture of
the responses in the different axes in the perpendicular plane. This must be
viewed as a drawback compared to the E-field method, where in principle the
response along each axis can be measured using a needle shaped specimen.

\subsubsection{Surface Impedance measurements (skin depth or eddy current
limit)}

The other useful limit is for a highly conducting material, where $%
ka=(1+i)a/\delta =(1+i)a\sqrt{\mu _{o}\omega \sigma }$. The skin depth $%
\delta $ $=1/\sqrt{\mu _{o}\sigma \omega }$ $\ll a$, hence 
\begin{equation}
\tilde{\zeta}_{H}-\zeta _{H\infty }^{\prime }=\frac{3}{\mu _{o}\omega a}%
(X_{s}+iR_{s})\,\,\,\,;\,\,\,\,\,\,\,\,\,\text{when\thinspace \thinspace }%
\,\,\,\tilde{\chi}_{M}\sim 0\text{, }a\gg \delta \text{, }ka\gg 1\text{ }
\end{equation}

It is useful to reference the data to the complete diamagnetic result $\zeta
_{H\infty }^{\prime }=-1.5$ for a sphere. Thus in this limit the data are a
direct measure of the surface impedance 
\begin{equation}
\tilde{Z}_{s}=R_{s}-iX_{s}=\sqrt{\frac{-i\omega \mu _{o}}{\tilde{\sigma}}}
\label{impedance}
\end{equation}
The normalization factor $3/(\mu _{o}\omega a)$ is specific to the spherical
sample and $TE_{011}$ mode geometries. Note that in this limit the measured
data are $\varpropto 1/\sqrt{\sigma }$.

It is worth noting that this result (Eq.\ref{impedance}) is also valid for
complex conductivity $\tilde{\sigma}=\sigma _{1}+i\sigma _{2}$ such as for a
superconductor. The above treatment assumes that displacement current
effects are negligible. If they are also present and can be represented in
terms of a dielectric constant $\tilde{\varepsilon}$, then we can also write 
\begin{equation}
\tilde{Z}_{s}=R_{s}-iX_{s}=\sqrt{\frac{-i\omega \mu _{o}}{\tilde{\sigma}%
-i\omega \tilde{\varepsilon}}}
\end{equation}

\subsubsection{Conductivity (Eddy Current) Peaks and Dielectric Loss Peaks}

\begin{figure}[tbph]
\begin{center}
\includegraphics*[width=0.4\textwidth]{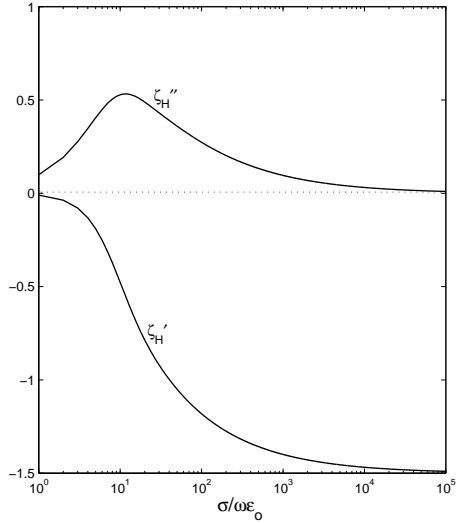}
\end{center}
\par
\caption{$\zeta _{H}^{\prime }$ and $\zeta _{H}^{\prime \prime }$ vs. $%
\sigma /i\omega \varepsilon _{0}$ showing the conductivity peak as $\sigma $
is varied.}
\label{eddycal}
\end{figure}

As noted above, the measured changes in the cavity resonance parameters
expressed here in terms of the electromagnetic susceptibility $\tilde{\zeta}%
_{H}$ change from a $\varpropto \sigma $ dependence for small $\sigma $ to a 
$\varpropto 1/\sqrt{\sigma }$ dependence for large $\sigma $. Thus as $%
\sigma $ is varied this results in a peak in the absorption or in $\zeta
_{H}^{\prime \prime }$, accompanied by a change of state of $\zeta
_{H}^{\prime }$ from $0$ to $\zeta _{H\infty }^{\prime }=-1.5$, as shown in
Fig.\ref{eddycal}. This conductivity or eddy current peak is similar to the
depolarization peak observed in $E$-field measurements. Of course the
location of the conductivity peak is determined by both the conductivity and
the sample dimensions.

In certain materials, particularly the oxides, there are dielectric loss
peaks intrinsic to the material, arising from a dielectric constant $\tilde{%
\varepsilon}=\varepsilon ^{\prime }+i\varepsilon ^{\prime \prime
}=\varepsilon (0)/(1+i\omega \tau )$. Usually $\tau $ is a strong function
of temperature $T$, and hence when $T$ is varied, a peak in $\varepsilon
^{\prime \prime }(T)$ occurs at a peak temperature $T_{p}$ where $\omega
\tau (T_{p})=1$. Since $\tau (T)$ increases with decreasing $T$, this peak
shifts to lower peak temperatures $T_{p}$ when the measurement frequency is
decreased. Since $\tilde{\zeta}_{H}\,$is proportional to $\tilde{\varepsilon}
$ in the appropriate limit, a peak will be observed in $\zeta _{H}^{\prime
\prime }$ also as $T$ is varied.

In such materials $\sigma (T)$ is a also a strong function of $T$ and
typically is semiconducting : $\sigma (T)=\sigma _{o}exp(-T_{s0}/T)$. Under
such conditions, the experimental data will display two peaks, one a
dielectric loss peak and the other a conductivity peak, as $T$ is varied.
When the measuring frequency $\omega $ is reduced, the dielectric loss peak
will move to lower $T$ while the conductivity peak will move to higher $T$,
i.e. the peaks move apart on the $T$ axis with decreasing $\omega $. A
specific example of a dielectric loss peak in the spin ladder material $%
Sr_{14}Cu_{24}O_{41}$ is discussed later.

\subsubsection{Dimensional Resonances}

\begin{figure}[tbph]
\begin{center}
\includegraphics*[width=0.5\textwidth]{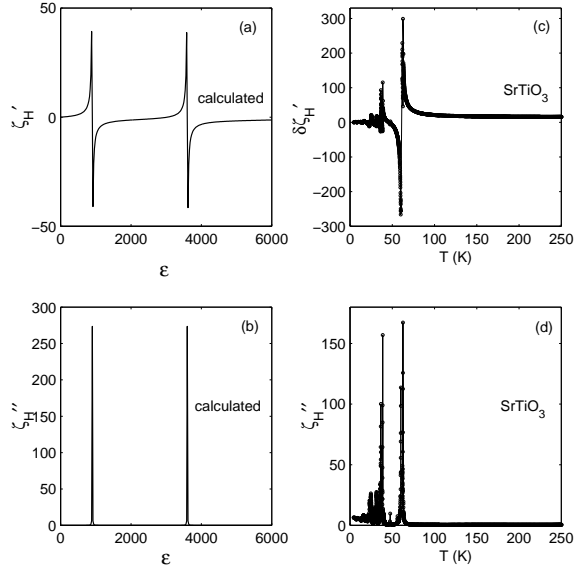}
\end{center}
\par
\caption{(a)$\zeta _{H}^{\prime }$ and (b) $\zeta _{H}^{\prime \prime }$ vs. 
$\varepsilon $ calculated using Eq. \ref{lossydielec}. The plots show the
dimensional resonances which occur when $ka=(n+1/2)\pi $. Experimental data
of (c) $\delta \zeta _{H}^{\prime }$ and (d) $\zeta _{H}^{\prime \prime }$
vs. $T$ for $SrTiO_{3}$. The dimensional resonances, similar to (a) and (b),
are clearly visible. In this material the dielectric constant $\varepsilon
^{\prime }$ increases with decreasing $T$. }
\label{dimexp}
\end{figure}

A remarkable prediction of Eq.\ref{lossydielec} is the occurrence of
dimensional resonances when the dielectric constant varies strongly. This is
shown in Fig.\ref{dimexp} (a) and (b). The resonances occur whenever $%
ka=(n+1/2)\pi $ and are quite sharp. They correspond to situations where the
electromagnetic field essentially resonates inside the sample, just like a
dielectric resonator. We have observed such resonances in $SrTiO_{3}$, which
can be viewed as a quantum paraelectric with transition temperature at $T=0$%
, and in which material $\varepsilon ^{\prime }$ increases rapidly with
decreasing $T$ to values approaching several thousands. Results are
discussed later.

\section{Experimental Procedures}

In the experiment, we first carry out a background run to measure the
resonance frequency $f_{c}(T)\,$and the width $\Delta f_{c}(T)$ of the empty
cavity as a function of $T$. Then the sample is inserted in and
corresponding parameters $f_{s}(T)$ and $\Delta f_{s}(T)$ are measured. $%
\tilde{\zeta}_{H}=\zeta _{H}^{\prime }(T)+i\zeta _{H}^{^{\prime \prime }}(T)$
is obtained using 
\begin{eqnarray}
\zeta _{H}^{\prime }(T) &=&-\frac{1}{g}(f_{s}(T)-f_{c}(T))
\label{zetaexpdef} \\
\zeta _{H}^{\prime \prime }(T) &=&\frac{1}{g}(\Delta f_{s}(T)-\Delta
f_{c}(T)).  \nonumber
\end{eqnarray}

$g$ is given by Eq. \ref{gforte011}. In practice, while relative changes $%
\delta f_{s}(T)=f_{s}(T)-f_{s}(T_{ref})$ or $\delta
f_{c}(T)=f_{c}(T)-f_{c}(T_{ref})$ referred to a reference temperature $%
T_{ref}$ can be measured with extremely high precision, there can be larger
errors in the absolute value of $f_{s}(T)-f_{c}(T)$. For this reason we
represent the data as 
\begin{equation}
\zeta _{H}^{\prime }(T)=-\frac{1}{g}\left[ (\delta f_{s}(T)-\delta
f_{c}(T))+\delta f(T_{ref})\right]  \label{abszetaprime}
\end{equation}

In many cases, the background correction $\delta f_{c}(T)$ can be
negligible. It is convenient to present the data as $\delta \zeta
_{H}^{\prime }(T)=\zeta _{H}^{\prime }(T)-\zeta _{H}^{\prime }(T_{ref})$
instead of $\zeta _{H}^{\prime }(T)$. To get the absolute value of $\zeta
_{H}^{\prime }$, calibration can be made by putting the sample into the
cavity to measure $f_{s}$ and then immediately taking the sample out to
measure $f_{c}$ at a fixed temperature, and thus obtain $\delta f(T_{ref})$.
For many samples, (e.g. see $Sr_{14}Cu_{24}O_{41}$ later), $\zeta
_{H}^{\prime }(T_{ref})\ll \zeta _{H}^{\prime }(T)$ particularly at high $T$%
, so that in these cases, $\delta \zeta _{H}^{\prime }\approx \zeta
_{H}^{\prime }$.

\subsection{Inversion of experimental $\tilde{\zeta}$ or $\tilde{Z}_{s}$
data to obtain $\tilde{\varepsilon}$ and $\tilde{\sigma}$}

The next key step is to obtain the fundamental material property, the sample
dielectric function $\tilde{\varepsilon}$ or conductivity $\tilde{\sigma}$,
from the experimental data represented either as $\tilde{\zeta}$ or the
surface impedance $\tilde{Z}_{s}$. Two approaches are possible here :

\begin{enumerate}
\item  A direct inversion of Eq.\ref{lossydielec} for $\tilde{\zeta}(T)$
data or Eq.\ref{impedance} for $\tilde{Z}_{s}(T)$ data to extract $\tilde{%
\sigma}(T)-i\omega \tilde{\varepsilon}(T)$,

\item  Modelling of $\tilde{\sigma}(T)-i\omega \tilde{\varepsilon}(T)$ to
quantitatively match the $\tilde{\zeta}(T)$ data using Eq.\ref{lossydielec}
or $\tilde{Z}_{s}(T)$ data using Eq.\ref{impedance}.
\end{enumerate}

We discuss both these procedures below.

\subsubsection{Inversion of equations}

We have successfully solved Eq.\ref{lossydielec} to obtain $\tilde{z}%
=ka=k_{o}a\sqrt{\tilde{\varepsilon}+i\tilde{\sigma}/\omega \varepsilon _{o}}$
using the subroutine {\em FSOLVE} in MATLAB. The success of the solution
depends crucially on the values of $\tilde{\zeta}$ or equivalently $\tilde{z}
$. For values of $\zeta ^{\prime },\zeta ^{\prime \prime }\sim 1$, which is
well in the QS or EQS limits, the solution is very accurate and yields the
sample $\tilde{\sigma}-i\omega \tilde{\varepsilon}$ with ease. In this limit 
$\tilde{z}\lesssim 1$, corresponding to typical dielectric constants $%
\varepsilon ^{\prime }\leq 1000$ (for the sample and cavity sizes discussed
in this paper) and not too small $\varepsilon ^{\prime \prime }$. Thus for
lossy dielectrics, the results for $\tilde{\varepsilon}$ can be easily
obtained. The results of such a solution for the material $%
Sr_{14}Cu_{24}O_{41}$ are discussed later in this paper. Results on several
other materials which have similar properties, such as $%
La_{5/3}Sr_{1/3}NiO_{4}$, $YBa_{2}Cu_{3}O_{6.0}$ and $\Pr
Ba_{2}Cu_{3}O_{7.0} $, are described in previous and forthcoming papers \cite
{Hakim99, Zhai99}.

Great care must be exercised in two regimes of parameter values :

\begin{enumerate}
\item  when $\zeta ^{\prime }\gg 1$, and $\zeta ^{\prime \prime }\leq 1$,
which corresponds to $ka\gg 1$ ( $\varepsilon ^{\prime }>1000$ for the
conditions of the experiments in this paper). Here the resonances of $\cot
(z)$ enter when $ka=(n+1)\pi /2$, leading to dimensional resonances
discussed in other sections.

\item  the metallic limit, when $\zeta ^{\prime }\rightarrow -1.5$, and $%
\zeta ^{\prime \prime }\ll 1$. In this limit it is more appropriate to use
the surface impedance limit Eq.\ref{impedance} rather than Eq.\ref
{lossydielec}.
\end{enumerate}

The principal difficulty in the above two limits is that there are many
nearby minima of the underlying function, and the program quickly converges
to spurious solutions. Future work will focus on this important problem.

\subsubsection{Modelling the conductivity and dielectric constant to match
the data}

Even if the solution procedure is successful and the material $\tilde{\sigma}%
-i\omega \tilde{\varepsilon}$ is successfully extracted, a quantitative
understanding of the experimental results for $\tilde{\sigma}-i\omega \tilde{%
\varepsilon}$ requires a model. Where the solution is not easily attained
due to the difficulties mentioned above, we have found it necessary to
bypass the solution procedure and instead use model calculations of $\tilde{%
\sigma}-i\omega \tilde{\varepsilon}$ to describe the $\tilde{\zeta}$ data
using Eq.\ref{lossydielec} and Eq.\ref{impedance}.

\section{Experimental Results}

We describe below measurements on three different materials all in single
crystal form. These crystals have typical dimensions of $1\times 1\times
0.5mm^{3}$ and have been extensively characterized by a vast array of
measurements: dc resistivity, dc SQUID susceptibility, XRD, neutron
scattering and high pressure studies. Structural studies of the single
crystals show that of all of these measurements indicate single phase, high
quality crystals.

\subsection{Magnetodynamics in the spin chain material $Sr_{2}CuO_{3}$}

\begin{figure}[tbph]
\begin{center}
\includegraphics*[width=0.45\textwidth]{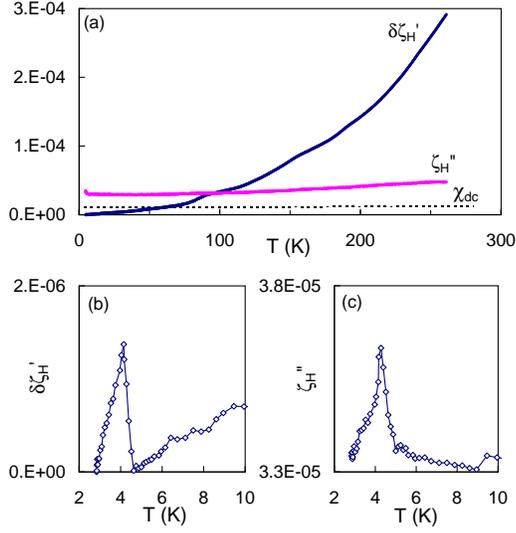}
\end{center}
\par
\caption{(a)Measured microwave magnetic susceptibility $\tilde{\zeta}_{H}$
vs. $T$ for $Sr_{2}CuO_{3}$. For this sample the limit $\tilde{\zeta}%
_{H}\approx \tilde{\chi}_{M}$ applies. Also shown is the dc susceptibility
(dashed line) from \protect\cite{Motoyama96}. (b) and (c) are low $T$ data
of $\delta \zeta _{H}^{\prime }$ and $\zeta _{H}^{\prime \prime }$, showing
signatures of the 3-D AFM transition at $T\thicksim 5K$. }
\label{sr2cuo3}
\end{figure}

$Sr_{2}CuO_{3}$ single crystals were prepared by the floating zone technique%
\cite{Vietkine}. It is an insulator in a large range of temperature and
there is only possesses linear $Cu-O$ chains and is regard as an ideal
one-dimensional spin $1/2$ chain. In the measurement, the sample is mounted
in such way that the microwave field $H_{\omega }//\hat{c}$-axis. Fig.\ref
{sr2cuo3} (a) shows the plot of $\delta \zeta _{H}^{\prime }(T)\equiv \zeta
_{H}^{\prime }(T)-\zeta _{H}^{\prime }(2K)$ {\em vs.} $T$ and $\zeta
_{H}^{\prime \prime }(T)$ {\em vs.} $T$ for $Sr_{2}CuO_{3}$. $\delta \zeta
_{H}^{\prime }$ shows a monotonic increase with temperature $T$, and $\zeta
_{H}^{\prime \prime }$ has insignificant changes from $6K$ to $260K$. These
results are consistent with the DC magnetic susceptibility measurements \cite
{Motoyama96}, as shown in dashed line of Fig.\ref{sr2cuo3} (a), and indicate
that the sample perturbation effect is in the magnetic susceptibility limit $%
(k_{o}a)^{2}\tilde{\chi}_{P}\ll \tilde{\chi}_{M}$, so that $\tilde{\zeta}%
_{H}\simeq \tilde{\chi}_{M}\,$. Thus in this material we are essentially
measuring the magnetic susceptibility.

At low temperatures, additional features are observed in both $\delta \zeta
_{H}^{\prime }$ and $\zeta _{H}^{\prime \prime }$, as shown in Fig.\ref
{sr2cuo3} (b) and (c). These peaks are microwave signatures of the 3D
Heisenberg AFM transition at $T_{N}\approx 5K$\cite{Motoyama96}. The very
high sensitivity of the technique utilizing a superconducting cavity is
evident from the data in Fig.\ref{sr2cuo3}.

\subsection{Dielectric Loss Peaks in the spin ladder material $%
Sr_{14}Cu_{24}O_{41}$}

\begin{figure}[tbph]
\begin{center}
\includegraphics*[width=0.55\textwidth]{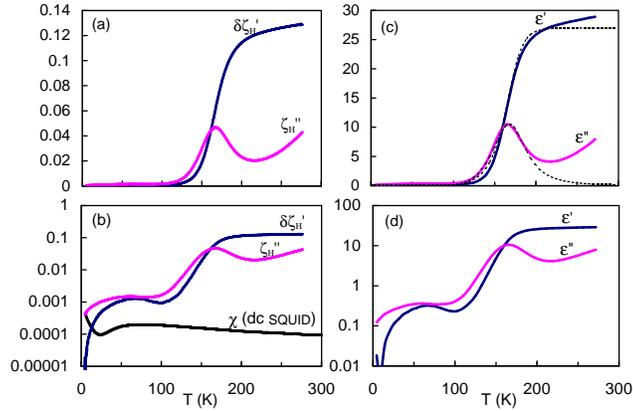}
\end{center}
\par
\caption{ Experimental data of $\delta \zeta _{H}^{\prime }$ ($\approx \zeta
_{H}^{^{\prime }}$) and $\zeta _{H}^{\prime \prime }$ for $%
Sr_{14}Cu_{24}O_{41}$ in (a) linear and (b) semilog plots. (c) $\varepsilon
^{\prime }$ and $\varepsilon ^{\prime \prime }$ obtained from the $\tilde{%
\zeta}_{H}$ data of (a) by solving Eq. \ref{lossydielec}. The dielectric
loss peak is clearly evident. Also shown in (c) (dashed line) is the fit of
the dielectric response to a Cole-Davidson form. (d) Data of (c) in a
semilog plot, showing an additional feature at $110K$ coincident with the
opening of the magnetic gap reported in this system. }
\label{sr14cu24o41}
\end{figure}

There is increasing interest for studying spin/ladder compounds because
superconductivity can be obtained in $Ca$ doped $Sr_{14}Cu_{24}O_{41}$ under
high pressures\cite{Uehara96}. In Fig. \ref{sr14cu24o41}, we show the
results of $\tilde{\zeta}(T)$ in the case of $H_{\omega }//\hat{c}$-axis for 
$Sr_{14}Cu_{24}O_{41}$. The striking feature of the data is the rapid drop
with decreasing $T$ in $\delta \zeta _{H}^{\prime \prime }(T)$ below
approximately $200K$, accompanied by a relatively sharp peak in $\zeta
_{H}^{\prime \prime }(T)$ at $T\sim 170K$, which is not seen in the DC
magnetic susceptibility measurement (Fig.\ref{sr14cu24o41} (b)). The
extraordinary dynamic range (over $4$ orders of magnitude in $\tilde{\zeta}%
_{H}$) of the superconducting cavity enables us to see an additional peak at
low $T$ in Fig.\ref{sr14cu24o41} (b) (the semilog plot of (a) data).
Although a similar peak is also observed in $\chi _{dc}(T)$, the magnitude
is about $10$ times smaller than $\zeta _{H}^{\prime \prime }(T)$. At high
temperatures, the measured $\zeta _{H}^{\prime }/\chi _{dc}\sim 10^{3}$.
Thus in this material the dielectric contributions dominate, {\em i.e. }$%
(k_{o}a)^{2}\tilde{\chi}_{P}\gg \tilde{\chi}_{M}$, and we are thus measuring
the dielectric constant.

Fig.\ref{sr14cu24o41} (c) shows the dielectric constant $\varepsilon
^{\prime }$ and $\varepsilon ^{\prime \prime }$ obtained from the measured $%
\tilde{\zeta}$ data and inverting Eq.\ref{lossydielec}. The loss peak in $%
\varepsilon ^{\prime \prime }$ is clearly evident, and is accompanied by a
change of state of $\varepsilon ^{\prime }$. These data indicate an
essentially pure dielectric relaxation process in this spin ladder material,
arising from the presence of charges due to doping.

The dielectric mode is well described by a Cole-Davidson form $\tilde{%
\varepsilon}(\omega ,T)=\varepsilon (0)/[1+i\omega \tau (T)]^{\beta }$, with 
$\varepsilon (0)=27$, $\beta =0.6$ and an activated relaxation time $\tau
(T)=1.6\times 10^{-16}\cdot \exp (T_{\tau 0}/T)$ [$\sec .]$, with an
activation energy $T_{\tau 0}=2000K$. When the relaxation rate $\tau
^{-1}(T) $ varies rapidly with $T$ and crosses the measurement frequency $%
\omega $, a peak occurs at $T_{p}$, $\,$where $\omega \tau (T_{p})=1$, as
shown in Fig.\ref{sr14cu24o41}. In this material the relaxation time $\tau
(T)$ appears to follow the conductivity $\sigma (T)$, indicating that the
free carriers determine the polarization relaxation. Extensive details of
the polarization dynamics in this material and in the related $%
Sr_{14-x}Ca_{x}Cu_{24}O_{41}$ family are discussed in a forthcoming
publication \cite{Zhai97}.

\subsection{Dimensional resonances in $SrTiO_{3}$}

One of the striking predictions of the above analysis is the occurrence of
dimensional resonances discussed in an earlier section. These resonances
occur when the dielectric constant is so large that the condition $%
ka=(n+1/2)\pi $ is satisfied. We have experimentally observed such
resonances in single crystal samples of $SrTiO_{3}$ measured in a $TE_{011}$
cavity. The single crystal samples were purchased from Aesar Mfg. Co.. The
experimental data are shown in Fig.\ref{dimexp} (c) and (d), where $\delta
\zeta _{H}^{\prime }(T)$ and $\zeta _{H}^{\prime \prime }(T)$ are shown as a
function of $T$ for a sample with dimensions $0.5\times 0.5\times 0.5mm^{3}$%
. The data clearly show resonances as a function of $T$. In this material $%
\varepsilon ^{\prime }$ increases strongly with decreasing $T$ approaching
values of nearly $1000$. The experimental data shown in Fig.\ref{dimexp} (c)
and (d) are quantitatively consistent with the behavior in (a) and (b).

Inversion of Eq.\ref{lossydielec} shows a weakly $T$-dependent $\varepsilon
^{\prime }\sim 850$ between $250K$ and about $75K$. This value is entirely
consistent with other measurements \cite{Ang99}. However at lower
temperatures the inversion of the $\tilde{\zeta}_{H}$ data to obtain $\tilde{%
\varepsilon}$ is problematical because of the dimensional resonances. Here
as noted before the solutions do not appear to be unique and spurious
solutions are found.

An important parameter for microwave applications is the microwave loss in $%
SrTiO_{3}$. We find that $\varepsilon ^{\prime \prime }(T)$ varies between $%
0.1-0.25$ in the temperature region between $250K$ and about $75K$ where
smooth solutions of $\varepsilon ^{\prime }\sim 850$ are obtained. We also
note that the raw experimental data for $\tilde{\zeta}$ indicate that the
biggest resonances occur at $62K$ and $37K$ which is exactly where
dielectric anomalies have been reported in lower temperature measurements 
\cite{Ang99}.

\subsection{Conclusion}

We have shown that a careful analysis of cavity perturbation methods,
combined with the use of superconducting cavities, leads to a powerful
method of measuring transport properties at microwave frequencies. The
method can lead to exceptionally high sensitivities for the material
properties.

A surprising result is that dielectric constants can be measured even though
the sample is placed in a microwave magnetic field. One consequence of this
conclusion is that many such experiments which use samples in microwave
magnetic fields, such as non-resonant microwave absorption measurements \cite
{MAMMA}, should be carefully analyzed for the influence of dielectric
properties, and not just the magnetic properties.

The resulting microwave measurements show new dynamic phenomena with time
scales corresponding to the GHz frequency ranges which is not seen in static
dc SQUID susceptibility measurements. The microwave measurements yield
information on dynamics at time scales $\sim 10^{-11}$ sec., comparable to
NS, but shorter than NMR and NQR ($10^{-7}$ sec.) and $\mu SR$ ($10^{-8}$
sec.), and are a sensitive probe of charge dynamics novel electronic
materials. These results will lead us to a new perspective of how to
understand other cuprates. In future publications, we will discuss the
results of measurements on low dimensional spin systems and high temperature
superconductors.

We thank A. Revcolevschi for providing samples of $Sr_{2}CuO_{3}$ and $%
Sr_{14}Cu_{24}O_{41}$, and R. S. Markiewicz for useful discussions. This
research was supported by NSF-9711910, AFOSR-F30602-95-2-0011 and
ONR-N00014-00-1-0002.

\pagebreak \appendix 

\section{Spherical sample in magnetic field maximum of TE$_{011}$ mode}

Brodwin and Parsons treated a spherical homogeneous sample with radius $a$
in a resonant cavity when the restrictions $ka\ll 1$ and $k_{o}a\ll 1$ are
removed. They use a method developed by Stratton\cite{Stratton} in which the
electric and magnetic fields inside and outside the perturbing sample are
expressed as expansions of spherical vector potential functions.

\strut The field configurations of the $TE_{011}$ mode in cylindrical
coordinate ($\hat{r},\hat{\varphi},\hat{z})$ are expressed as 
\begin{eqnarray}
\vec{H}(r,\varphi ,z) &=&-H_{o}\frac{\pi }{\beta _{01}^{^{\prime }}L}%
J_{1}(\beta _{01}^{^{\prime }}r)\cos (\frac{\pi z}{L})\,\hat{r}%
+H_{o}J_{0}(\beta _{01}^{^{\prime }}r)\sin (\frac{\pi z}{L})\,\hat{z}
\label{te011fields} \\
\vec{E}(r,\varphi ,z) &=&H_{o}\frac{i\omega \mu _{o}}{\beta _{01}^{^{\prime
}}}J_{1}(\beta _{01}^{^{\prime }}r)\sin (\frac{\pi z}{L})\,\hat{\varphi} 
\nonumber
\end{eqnarray}
where $H_{o}$ is the maximum magnetic field in the center of the cavity, $%
r_{o}$ is the radius of the cavity, $L$ is the cavity axial length, and $%
\beta _{01}^{^{\prime }}r_{o}$ is the first root of the Bessel function $%
J_{0}^{\prime }(\beta r_{o})=0$. A time dependence $e^{-i\omega t}$ is
assumed in $\vec{H}$ and $\vec{E}$. When a small sample with radius $a$ is
put inside the cavity, the complex frequency shift $\delta \tilde{\omega}$
for $TE_{011}$ mode is given by the following expression\cite{Brodwin65} 
\begin{equation}
\frac{\delta \tilde{\omega}}{\omega }=\frac{i9\eta \sin ^{2}\alpha }{%
2J_{0}^{2}(\beta _{01}^{\prime }r_{o})}\sum\limits_{n=1}^{\infty }\frac{%
2(2n+1)}{3n(n+1)}\left[ \frac{P_{n}^{\prime }(\cos \alpha )}{\sin \alpha }%
\right] ^{2}\delta _{0n}\left( \frac{a_{n}^{r}}{\rho ^{3}}\right)
\label{bpgeneq1}
\end{equation}
where $\beta =\sqrt{k_{o}^{2}-h^{2}}=\kappa _{o}\sin \alpha $, $h=\pi /L$,
and $\eta =V_{s}/V_{c}$ is the filling factor. $a_{n}^{r}$ is the
coefficient corresponding to the reflected (scattered) field and is given by
the following expression with $\rho =k_{o}a$ and $N\rho =ka$\cite{Brodwin65}%
. 
\begin{equation}
a_{n}^{r}=\frac{-\tilde{\mu}j_{n}(N\rho )[\rho j_{n}(\rho )]^{^{\prime
}}+j_{n}(\rho )[N\rho j_{n}(N\rho )]^{^{\prime }}}{\tilde{\mu}j_{n}(N\rho
)[\rho h_{n}^{1}(\rho )]^{^{\prime }}-h_{n}^{1}(\rho )[N\rho j_{n}(N\rho
)]^{^{\prime }}}  \label{bpcoeff1}
\end{equation}

In the following we define the sample geometrical factor $\gamma $ as 
\begin{equation}
\gamma =\frac{\eta \sin ^{2}\alpha }{J_{0}^{2}(\beta _{01}^{\prime }r_{o})}
\label{gamdef}
\end{equation}
Considering the cavity resonant frequency $\omega _{mnp}=c\sqrt{\beta
_{mn}^{^{\prime }2}+p^{2}\pi ^{2}/L^{2}}$, $\gamma $ for the $TE_{011}$ mode
can be rewritten as 
\begin{equation}
\gamma =\frac{\eta }{J_{0}^{2}(\beta _{01}^{\prime }r_{o})\left[ 1+\left( 
\frac{\pi }{\beta _{01}^{\prime }L}\right) ^{2}\right] }  \label{gamdef2}
\end{equation}

This series in Eq.\ref{bpgeneq1} is rapidly convergent for samples with
diameters less than $\lambda /2\pi ,$ therefore the leading term give a good
approximation for the frequency shift. 
\begin{equation}
\frac{\delta \tilde{\omega}}{\omega }=\frac{i9\gamma a_{1}^{r}}{2\rho ^{3}}
\label{bpgeneq3}
\end{equation}

Using the spherical Bessel functions: $j_{1}(\rho )=[\sin (\rho )-\rho \cos
(\rho )]/\rho ^{2}$, $h_{1}^{(1)}(\rho )=-e^{i\rho }(\rho +i)/\rho ^{2}$,
and $[\rho j_{1}(\rho )]^{^{\prime }}=-[\sin (\rho )-\rho \cos (\rho )]/\rho
^{2}+\sin (\rho )$, we can examine the results of $\delta \tilde{\omega}$ in
various limits.

\subsection{Extended Quasistatic Limit $k_{o}a\ll 1$}

In this approximation, $j_{1}(\rho )=\rho /3-\rho ^{3}/30$, $%
h_{1}^{(1)}(\rho )=-i/\rho ^{2}-i/2+\rho /3+O(3)$, and $[\rho j_{1}(\rho
)]^{^{\prime }}=2\rho /3-2\rho ^{3}/15$, $[\rho h_{1}^{(1)}(\rho
)]^{^{\prime }}=i/\rho ^{2}-i/2+2\rho /3+O(3)$, Eq.\ref{bpgeneq1} can be
written as 
\begin{equation}
\frac{\delta \tilde{\omega}}{\omega }=-\frac{3\gamma }{2}\left[ \frac{2%
\tilde{\mu}j_{1}(N\rho )-[N\rho j_{1}(N\rho )]^{^{\prime }}}{\tilde{\mu}%
j_{1}(N\rho )+[N\rho j_{1}(N\rho )]^{^{\prime }}}\right] +A_{1}\cdot
(k_{o}a)^{2}  \label{bpgeneq2}
\end{equation}
where $A_{1}$ is coefficient of the second order term of $k_{o}a$ and given
by: 
\begin{eqnarray}
A_{1} &=&\frac{3\gamma }{20}\left[ \frac{4\tilde{\mu}j_{1}(N\rho )-\left[
N\rho j_{1}(N\rho )\right] ^{^{\prime }}}{\tilde{\mu}j_{1}(N\rho )+\left[
N\rho j_{1}(N\rho )\right] ^{^{\prime }}}\right]  \label{koa2th} \\
&&-\frac{3\gamma }{4}\left[ \frac{\left( 2\tilde{\mu}j_{1}(N\rho )-\left[
N\rho j_{1}(N\rho )\right] ^{^{\prime }}\right) \left( \tilde{\mu}%
j_{1}(N\rho )-\left[ N\rho j_{1}(N\rho )\right] ^{^{\prime }}\right) }{%
\left( \tilde{\mu}j_{1}(N\rho )+\left[ N\rho j_{1}(N\rho )\right] ^{^{\prime
}}\right) ^{2}}\right]  \nonumber
\end{eqnarray}

\subsubsection{Quasistatic Limit $k_{o}a\ll 1$}

Considering $N\rho =ka\ll 1$ the frequency shift $\delta \tilde{\omega}$ can
be reduced to: 
\begin{equation}
\frac{\delta \tilde{\omega}}{\omega }=-3\gamma \frac{\tilde{\mu}-1}{\tilde{%
\mu}+2}-\frac{9\gamma }{10}\left[ \frac{\tilde{\mu}^{2}-6\tilde{\mu}+4}{%
\left( \tilde{\mu}+2\right) ^{2}}\left( k_{o}a\right) ^{2}+\frac{\tilde{\mu}%
}{\left( \tilde{\mu}+2\right) ^{2}}\left( ka\right) ^{2}\right]
\label{bpmaglimit}
\end{equation}

If $\tilde{\mu}=1+\tilde{\chi}_{m}\approx 1$, in addition to $k_{o}a$, $%
ka\ll 1$, then the above equation reduces to 
\begin{eqnarray}
\frac{\delta \tilde{\omega}}{\omega } &\approx &-\gamma (\widetilde{\mu }-1)
\label{bpmaglimit2} \\
&=&-\gamma \tilde{\chi}_{M}=-\gamma (\chi _{M}^{^{\prime }}+i\chi
_{M}^{^{\prime \prime }})  \nonumber
\end{eqnarray}

Here the frequency shift $\delta \tilde{\omega}$ is a measurement of the
complex magnetic susceptibility $\tilde{\chi}_{M}$.

\subsubsection{Pure conductor: Eddy current or skin depth limit: $\tilde{\mu}%
=1$, $\tilde{\varepsilon}=1$, $\tilde{\sigma}=\sigma $}

In this limit $ka=(1+i)a/\delta ,$ where $\delta $ $=1/\sqrt{\mu _{o}\tilde{%
\sigma}\omega }$ is the skin depth. Retaining the first order in the series
of Eq.\ref{bpgeneq2}, we obtain the complex frequency shift $\delta \tilde{%
\omega}$: 
\begin{eqnarray}
\frac{\delta \tilde{\omega}}{\omega } &=&-\frac{3\gamma }{2}\left[ \frac{%
2j_{1}(N\rho )-[N\rho j_{1}(N\rho )]^{^{\prime }}}{j_{1}(N\rho )+[N\rho
j_{1}(N\rho )]^{^{\prime }}}\right]  \label{bpeddy1} \\
&\approx &\frac{3\gamma }{2}\left[ 1-\frac{3}{(ka)^{2}}+\frac{3\cot ka}{ka}%
\right]  \nonumber
\end{eqnarray}

By using $\cot (x+iy)=\sin 2x/(\cosh 2y-\cos 2x)-i\sinh 2y/(\cosh 2y-\cos
2x) $, we obtain the following expressions: 
\begin{eqnarray}
\func{Re}\left( \frac{\delta \tilde{\omega}}{\omega }\right) &=&\frac{%
3\gamma }{2}\left[ 1-\frac{3}{2}\left( \frac{\delta }{a}\right) \left( \frac{%
\sinh \frac{2\delta }{a}-\sin \frac{2\delta }{a}}{\cosh \frac{2\delta }{a}%
-\cos \frac{2\delta }{a}}\right) \right]  \label{bpeddy2} \\
\func{Im}\left( \frac{\delta \tilde{\omega}}{\omega }\right) &=&\frac{%
9\gamma }{4}\left( \frac{\delta }{a}\right) ^{2}\left[ 1-\left( \frac{a}{%
\delta }\right) \left( \frac{\sinh \frac{2\delta }{a}+\sin \frac{2\delta }{a}%
}{\cosh \frac{2\delta }{a}-\cos \frac{2\delta }{a}}\right) \right]  \nonumber
\end{eqnarray}

In the low frequency limit where $\delta \gg a$ the above formulas become: 
\begin{eqnarray}
\func{Re}\left( \frac{\delta \tilde{\omega}}{\omega }\right) &=&\frac{%
4\gamma }{105}\left( \frac{\delta }{a}\right) ^{4}  \label{bpeddy3} \\
\func{Im}\left( \frac{\delta \tilde{\omega}}{\omega }\right) &=&\frac{\gamma 
}{5}\left( \frac{\delta }{a}\right) ^{2}  \nonumber
\end{eqnarray}

In the high frequency limit where $\delta \ll a$ we obtain the expressions: 
\begin{eqnarray}
\func{Re}\left( \frac{\delta \tilde{\omega}}{\omega }\right) &=&\frac{%
3\gamma }{2}-\frac{9\gamma }{4}\left( \frac{\delta }{a}\right)
\label{bpeddy4} \\
\func{Im}\left( \frac{\delta \tilde{\omega}}{\omega }\right) &=&\frac{%
9\gamma }{4}\left( \frac{\delta }{a}\right)  \nonumber
\end{eqnarray}

Therefore the complex frequency shift $\delta \tilde{\omega}$ could be
written in terms of surface impedance $Z_{s}=R_{s}-iX_{s}$: 
\begin{equation}
\frac{\delta \tilde{\omega}}{\omega }=\frac{3\gamma }{2}\left[ 1-\frac{3}{%
\omega \mu _{o}a}\left( X_{s}+iR_{s}\right) \right]  \label{bpsurfimp}
\end{equation}
with $R_{s}=X_{s}=\sqrt{\omega \mu _{o}/2\sigma }$.

\subsubsection{Lossy Dielectric : $\tilde{\mu}=1,$ $\tilde{\varepsilon}%
=\varepsilon ^{\prime }+i\varepsilon ^{\prime \prime }$, $\tilde{\sigma}%
=\sigma $.}

In this case, $k^{2}=(\omega /c)^{2}(\tilde{\varepsilon}+i\tilde{\sigma}%
/\omega \varepsilon _{o})$, the frequency shift has a similar expression
with the one derived for a perfect conductor but in this case the real and
the imaginary part of the wave vector are not equal. 
\begin{equation}
\frac{\delta \tilde{\omega}}{\omega }=\frac{3\gamma }{2}\left[ 1-\frac{3}{%
(ka)^{2}}+\frac{3\cot ka}{ka}\right]  \label{bplossy1}
\end{equation}

In the limit where $ka\ll 1$, the above equation can be written as 
\begin{eqnarray}
\frac{\delta \tilde{\omega}}{\omega } &\approx &-\frac{\gamma }{10}%
(k_{o}a)^{2}(\tilde{\varepsilon}+i\frac{\tilde{\sigma}}{\omega \varepsilon
_{o}}-1)  \label{bplossy2} \\
&\approx &-\frac{\gamma }{10}(k_{o}a)^{2}\tilde{\chi}_{P}\text{ ;
\,\,\,\,\,(when \,\,\,\,\,}\tilde{\sigma}=0\text{)}  \nonumber
\end{eqnarray}
where $\tilde{\chi}_{P}\equiv \tilde{\varepsilon}-1=\varepsilon ^{\prime
}-1+i\varepsilon ^{\prime \prime }=\chi _{P}^{\prime }+i\chi _{P}^{\prime
\prime }$. Here the frequency shift $\delta \tilde{\omega}$ is a measurement
of the complex dielectric susceptibility $\tilde{\chi}_{P}$ when $\tilde{%
\sigma}=0$.

\section{Sample in TM$_{110}$ Electric Field Maximum}

Although we have focussed on the $TE_{011}$ mode, it is also possible to
carry out measurements using the $TM_{110}$ mode. For a sample placed in the
cavity center at the microwave electric field maximum, the frequency shift
is \cite{Brodwin65}: 
\begin{equation}
\frac{\delta \tilde{\omega}}{\omega }=\frac{i9\eta }{4J_{1}^{2}(\beta
_{01}^{\prime }r_{o})}\sum\limits_{n=1}^{\infty }\frac{2(2n+1)}{3n(n+1)}%
\left[ P_{n}^{\prime }(0)\right] ^{2}\delta _{0n}\left( \frac{b_{n}^{r}}{%
\rho ^{3}}\right)  \label{bptmeq1}
\end{equation}
with the reflection coefficient \cite{Brodwin65}: 
\begin{equation}
b_{n}^{r}=\frac{-\tilde{\varepsilon}j_{n}(N\rho )[\rho j_{n}(\rho
)]^{^{\prime }}+j_{n}(\rho )[N\rho j_{n}(N\rho )]^{^{\prime }}}{\tilde{%
\varepsilon}j_{n}(N\rho )[\rho h_{n}^{1}(\rho )]^{^{\prime }}-h_{n}^{1}(\rho
)[N\rho j_{n}(N\rho )]^{^{\prime }}}  \label{bptmeq2}
\end{equation}

In the first order the Eq.\ref{bptmeq1} becomes: 
\begin{equation}
\frac{\delta \tilde{\omega}}{\omega }=-\frac{3\gamma ^{^{\prime }}}{2}\left( 
\frac{2\tilde{\varepsilon}j_{1}(N\rho )-[N\rho j_{1}(N\rho )]^{^{\prime }}}{%
\tilde{\varepsilon}j_{1}(N\rho )+[N\rho j_{1}(N\rho )]^{^{\prime }}}\right)
\label{bptmeq3}
\end{equation}
with a new geometrical factor $\gamma ^{^{\prime }}$ given by: 
\begin{equation}
\gamma ^{^{\prime }}=\frac{\eta }{2J_{1}^{2}(\kappa _{01}r_{o})}
\label{bptmeq4}
\end{equation}
where $\kappa _{01}r_{o}$ is the first root of the Bessel function $%
J_{0}(\kappa r_{o})=0$

In the limit where $ka\ll 1$ the frequency shift is: 
\begin{equation}
\frac{\delta \tilde{\omega}}{\omega }=-3\gamma ^{^{\prime }}\frac{\tilde{%
\varepsilon}-1}{\tilde{\varepsilon}+2}+O(2)
\end{equation}
\newline

\end{document}